**SciencePG**
Science Publishing Group

# The Application of Data Mining in the Production Processes

**Hamza Saad**

System Science and Industrial Engineering, Binghamton University, New York, USA

**Email address:**
Hsaad1@binghamton.edu

**Abstract:** Traditional statistical and measurements are unable to solve all industrial data in the right way and appropriate time. Open markets mean the customers are increased, and production must increase to provide all customer requirements. Nowadays, large data generated daily from different production processes and traditional statistical or limited measurements are not enough to handle all daily data. Improve production and quality need to analyze data and extract the important information about the process how to improve. Data mining applied successfully in the industrial processes and some algorithms such as mining association rules, and decision tree recorded high professional results in different industrial and production fields. The study applied seven algorithms to analyze production data and extract the best result and algorithm in the industry field. KNN, Tree, SVM, Random Forests, ANN, Naïve Bayes, and AdaBoost applied to classify data based on three attributes without neglect any variables whether this variable is numerical or categorical. The best results of accuracy and area under the curve (ROC) obtained from Decision tree and its ensemble algorithms (Random Forest and AdaBoost). Thus, a decision tree is an appropriate algorithm to handle manufacturing and production data especially this algorithm can handle numerical and categorical data.

## 1. Introduction

In manufacturing and industrial processes, large data generated daily, however traditional measurements and analysis are not enough to deal with the huge data. The problem comes from complexity, non-linear, and inconsistent. With the fast growth of the databases in the numerous modern organization, data mining becomes an increasingly valuable data analysis method. The application of databases and statistical analysis are well established in the engineering [2]. The first using of artificial intelligence in the manufacturing was developed in the late 1980s [3-4]. Data from all the processes of the companies such as product design, materials control, and planning, scheduling, assembly, recycling, maintenance, are recorded. These data are stored therefore offer big potential as sources of new knowledge. Making utilize of the data collected is become an issue, and data mining is a typical and best solution for transforming these data into knowledge.

The application of data mining process in manufacturing started about the 1990s [5-7], and it is gradually developed by receiving the attention from the production communities. Currently, data mining applied to different areas in production and manufacturing engineering to extract knowledge to apply in predicting maintenance, design, fault detection, quality control, production, and decision support systems. Data analyzed by data mining to identify the hidden patterns in the variables which control the manufacturing processes or to define and improve the quality of products. A significant advantage of data mining is that the data that required for analysis can be collected during normal processes and operations of the manufacturing process being studied and it is therefore generally are not necessary to introduce the dedicated processes for the data collection. Data mining in manufacturing has increased over the last years; it is currently suitable to review its application critically [10].

Due to complexity in manufacturing, data mining offers many algorithms to deal with the complicated data. Each algorithm has a particular ability to deal with categorical or numerical data. In this study, many data mining algorithms applied to classify data based on three attributes. In industry and manufacturing processes, association rules and decision tree are appropriated to extract the knowledge from big data. However, association rules can only handle categorical data. Decision tree and ensemble learning based decision tree recorded sophisticated output [1] whether the dependent variable is numerical or categorical. The algorithms of data mining in this study used to analyze data in [1] which originally obtained from an industrial progress report, and figure out the best algorithm can handle industrial data with the highest classification and performance.

# 2. Methodology (Overview of Data Mining Algorithms)

All algorithms of data mining which explained will be applied for analyzing the real problem in the production process.

## 2.1. Naive Bayes

The classifier of Naive Bayesian is generated from Bayes' theorem with the independence assumptions between variables. A Naive Bayesian classifier is easy to build, with no problematic iterative parameter estimation which makes it useful for huge industrial datasets. Regardless of its simplicity, the algorithm often does surprisingly well and is widely applied because it often outperforms perfect classification methods.

Bayes theorem in equation (1) is providing a technique of calculating the posterior probability, P(c|x), from P(c), P(x), and P(x|c). It assumes that the effect of the predictor (x) on a given class (c) is independent of the values of other input variables. This assumption is named the class conditional independence.

$$P(c/x) = \frac{P(x/c) * P(c)}{P(x)} \qquad (1)$$

P(c/x) is a posterior probability of class (dependent) given predictor (variable).

P(x/c) is a likelihood which is the probability of predictor given class.

P(c) is a prior probability of class.

P(x) is a prior probability of predictor [9].

## 2.2. Decision Tree Algorithm (C4.5)

A decision tree is conducting regression and classification models in the form of a tree building. It is breaking down the dataset into smaller subsets, and simultaneously the associated decision tree is developed gradually [11]. The ultimate result is a tree with leaf nodes and decision nodes. A decision node (e.g., machine) has two or more branches (e.g., old, Overcast and new). Leaf node (e.g., average) represents a decision. The root node of the tree is at the topmost decision node in a tree which corresponds to the best predictor. Decision trees can handle both numerical and categorical data.

The base of the algorithm for building decision trees named ID3 by J. R. Quinlan which conducts a top-down, greedy search through the possible branches space of with no backtracking. ID3 applies Information Gain and Entropy to construct the decision tree.

## 2.3. K-Nearest Neighbor Algorithm (KNN)

It is a simple algorithm that stored all available cases and classified the new cases based on the similarity measure (e.g., distance functions). K-Nearest Neighbor has used as a non-parametric technique in the pattern recognition and statistical estimation at the beginning of the 1970's [9].

Euclidean

$$\sqrt{\sum_{i=1}^{k}(x_i - y_i)^2} \qquad (2)$$

Manhatta

$$\sum_{i=1}^{k}|x_i - y_i| \qquad (3)$$

Minkowski

$$\left(\sum_{i=1}^{k}(|x_i - y_i|)^q\right)^{1/q} \qquad (4)$$

The three distance measure functions are only applying for continuous data. In the categorical instance variables, the Hamming distance in equation (5) must use instead of the distance function. The standardization of the numerical variables is between 0 and 1 when there is a mixture between numerical and categorical variables in the dataset.

Hamming Distance

$$D_H = \sum_{i=1}^{k}|x_i - y_i|$$

$$x = y \Rightarrow D = 0$$

$$x \neq y \Rightarrow D = 1 \qquad (5)$$

Choosing the optimal value for K is best done by first inspecting the data. In general, a large K value is more precise as it reduces the overall noise, but there is no guarantee. Cross-validation is another way to retrospectively determine a good K value by using an independent dataset to validate the K value. Historically, the optimal K for most datasets has been between 3-10. That produces much better results than 1NN.

## 2.4. Artificial Neural Networks Algorithm (ANN)

An artificial neural network (ANN) is a system that is based on biological neural networks, like a brain. An ANN is comprised of an artificial neurons network (which known as "nodes"). These nodes connected to each other in network shape, and a strength of the connections to another is assigned in the value based on strength: the inhibition (maximum being -1 and 0) or the excitation (maximum being +1 and 0). If a value of the connection is high, that is indicating that there is a strong connection. Within each design of the node, the transfer function is calculated. Three types of neurons in an Artificial Neural Network are input node, hidden node, and the output node. The input nodes take in the information, in the form of which can be explained numerically. The information presented the activation values, where each node gives a number, the higher the number means the huge activation. This information is then passed throughout the whole network. Based on the connection weights (strengths), transfer functions, and excitation or inhibition, the activation value is passed through the node to node. Each of the nodes sums the activation values it receives; it is then modifying the value based on its transfer function. The activation flowed through the network, through the hidden layer, until it reached the output nodes. Then, the output nodes reflect the input in a meaningful way to an outside world [Saad's dissertation 2018].

### 2.5. Support Vector Machine Algorithm (SVM)

A Support Vector Machine (SVM) performs classification by finding the hyperplane that maximizes the margin between the two classes. The vectors (cases) that define the hyperplane are the support vectors [11].

Algorithm defines an optimal hyperplane: maximize margin.

Extend the above definition for non-linearly separable problems: have a penalty term for misclassifications.

Map data to high dimensional space where it is easier to classify with linear decision surfaces: reformulate problem so that data is mapped implicitly to this space.

To define an optimal hyperplane, we need to maximize the width of the margin (w). SVM uses equation (6-8), and according to the type of data [9] that need to solve.

Linear SVM

$$x_i . x_j \qquad (6)$$

Non-linear SVM

$$\emptyset(x_i) . \emptyset \qquad (7)$$

Kernel Function

$$k (x_{(i)} . x_{(j)}) \qquad (8)$$

### 2.6. AdaBoost Algorithm

The AdaBoost (short for "Adaptive boosting") widget is a machine-learning algorithm, formulated by (Yoav Freund) and (Robert Schapire). It can be used with other learning algorithms to boost their performance. It does so by tweaking the weak learners.

An ensemble meta-algorithm that combines weak learners and adapts to the 'hardness' of each training sample. It works for both classification and regression.

### 2.7. Random Forests Tree Algorithm

It is an ensemble learning technique applied for regression, and classification. It was first invented by (Tin Kam Ho) and then it developed by (Breiman, 2001).

The algorithm is building many decision trees according to the user request. The bootstrap sample used to develop each tree based on the training data. When one tree developed, the arbitrary subset of attributes is conducted, from which the best attribute for a split is voted. The final result goes to two calculation. If the dependent variable is numerical, then the final result will be based on the average of all results. If the dependent variable is categorical, then the final result is based on the majority vote from individually developed trees in the forest. It merely works for both regression and classification tasks.

Some algorithms can handle categorical and numerical data, and other algorithms can handle only categorical data such as Naïve Bayes. Data for this study is included categorical and numerical variables to test the ability of each algorithm in handling and analyzing data.

## 3. Application of Data Mining Algorithms to Solve a Production Problem

Data is applied and fully explained in [1]. Data includes 12 input variables (Badge No, Job Title, Base Production, Production Achieved, Incentive Wages, Production Rate, Worker Efficiency, Machine Model, Product Type, Elapsed Time, and Unit of Production) and one output variable is (Ultimate Performance Evaluation), each variable in the dataset has 121 instances. It is collected from daily performance measurement of the worker based on the final evaluation for each shift. Algorithms applied to classify data based on the output of Ultimate Performance Evaluation. All algorithms in the study have the ability for classification and regression, but data modified in classification model because we need to measure the ability of all algorithms in the same solution application (Included Naïve Bayes which is categorical) and compare each one with the other algorithms in this study.

In figure 1, Orange software applied to analyze data because this software has perfect presentation and explanation to the results.

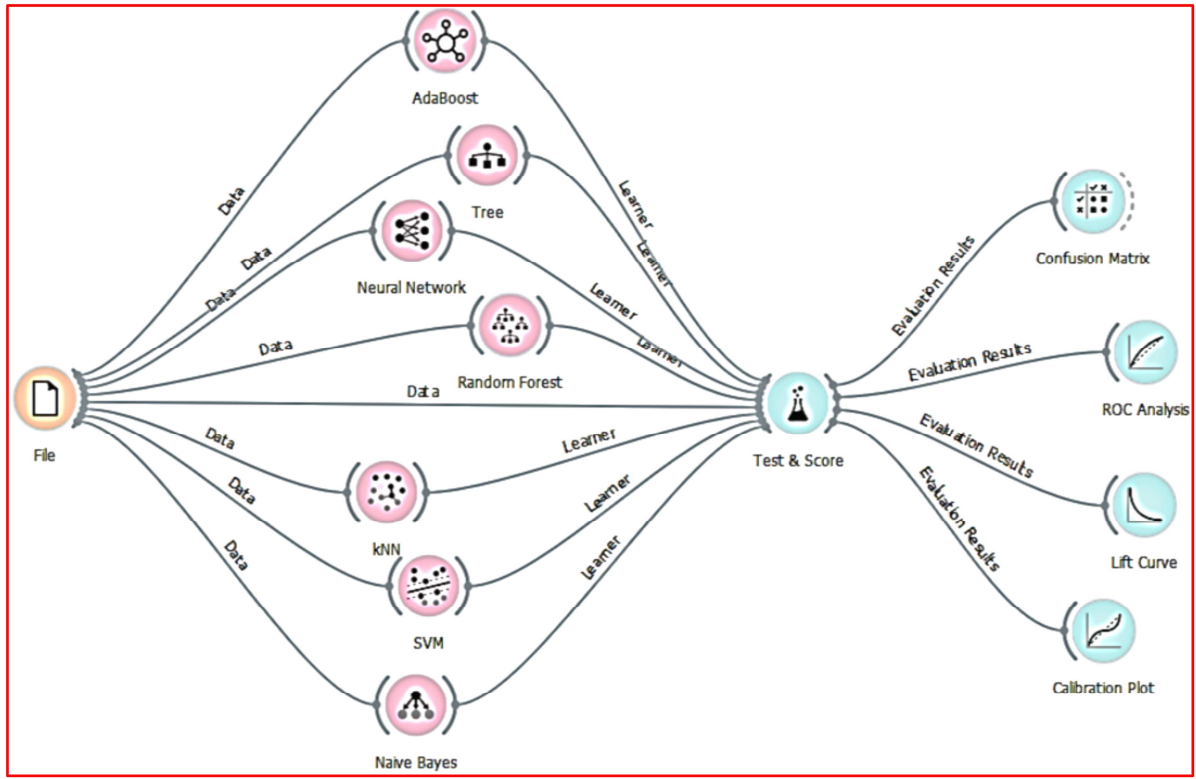

***Figure 1.*** *Apply different data algorithms using orange software.*

### 3.1. Results and Discussion

In classification accuracy, the best accuracy came from AdaBoost and Decision tree, then Random Forests and Naïve Bayes respectively. In previous studies [8], association rules and decision tree tested in the industry field, and they approved that the algorithms are appropriate to handle manufacturing data. However, association rules cannot handle numerical data or solve in regression-type especially huge daily data collected from industry is numerical. Decision tree and its ensemble learning algorithms such as

Random Forests and AdaBoost (Boosted tree) can handle and analyze data in regression or classification with the high efficiency. In the table, decision tree and its ensemble algorithms (AdaBoost and Random Forests) gave the highest classification accuracy beside Naïve Bayes which gave accuracy (0.826). K Nearest Neighbor (KNN) gave the lowest Classification Accuracy. Thus, we can support a decision about Decision tree as the best algorithm to deal with non-linear and manufacturing data.

***Table 1.*** *Outlines the results which obtained by applying seven algorithms of data mining.*

| Algorithm | AUC (Area Under ROC Curve | CA (Classification accuracy) | F1 | Precision | Recall |
|---|---|---|---|---|---|
| KNN | 0.668 | 0.636 | 0.662 | 0.652 | 0.672 |
| Tree | 0.879 | 0.868 | 0.873 | 0.887 | 0.859 |
| SVM | 0.721 | 0.769 | 0.800 | 0.737 | 0.875 |
| Random Forests | 0.932 | 0.860 | 0.870 | 0.851 | 0.891 |
| ANN | 0.788 | 0.702 | 0.727 | 0.706 | 0.750 |
| Naïve Bayes | 0.908 | 0.826 | 0.821 | 0.906 | 0.750 |
| AdaBoost | 0.868 | 0.868 | 0.875 | 0.875 | 0.875 |

Table 1. Performance of each algorithm of data mining.

### 3.2. Confusion Matrix Obtained from Each Algorithm

In the output variable of the dataset, there are three attributes or classes (Average, Good, and Excellent), confusion matrix can explain how the algorithms work based on each attribute. (Average instance), the attribute is highly classified using Tree, and Naïve Bayes with 22 of 27 instances and the lowest classification came from support vector machine with 15 of 27 instances. (Excellent instance), the attribute is highly classified using AdaBoost with 27 of 30 instances, and the lowest classification came from K Nearest Neighbor (KNN) with 11 of 30 instances. (Good instance), the attribute is highly classified using support vector machine (SVM), and AdaBoost with 56 of 64 instances and the lowest classification came from K Nearest Neighbor (KNN) with 43 of 46 instances. The confusion matrix is presented in figure 2.

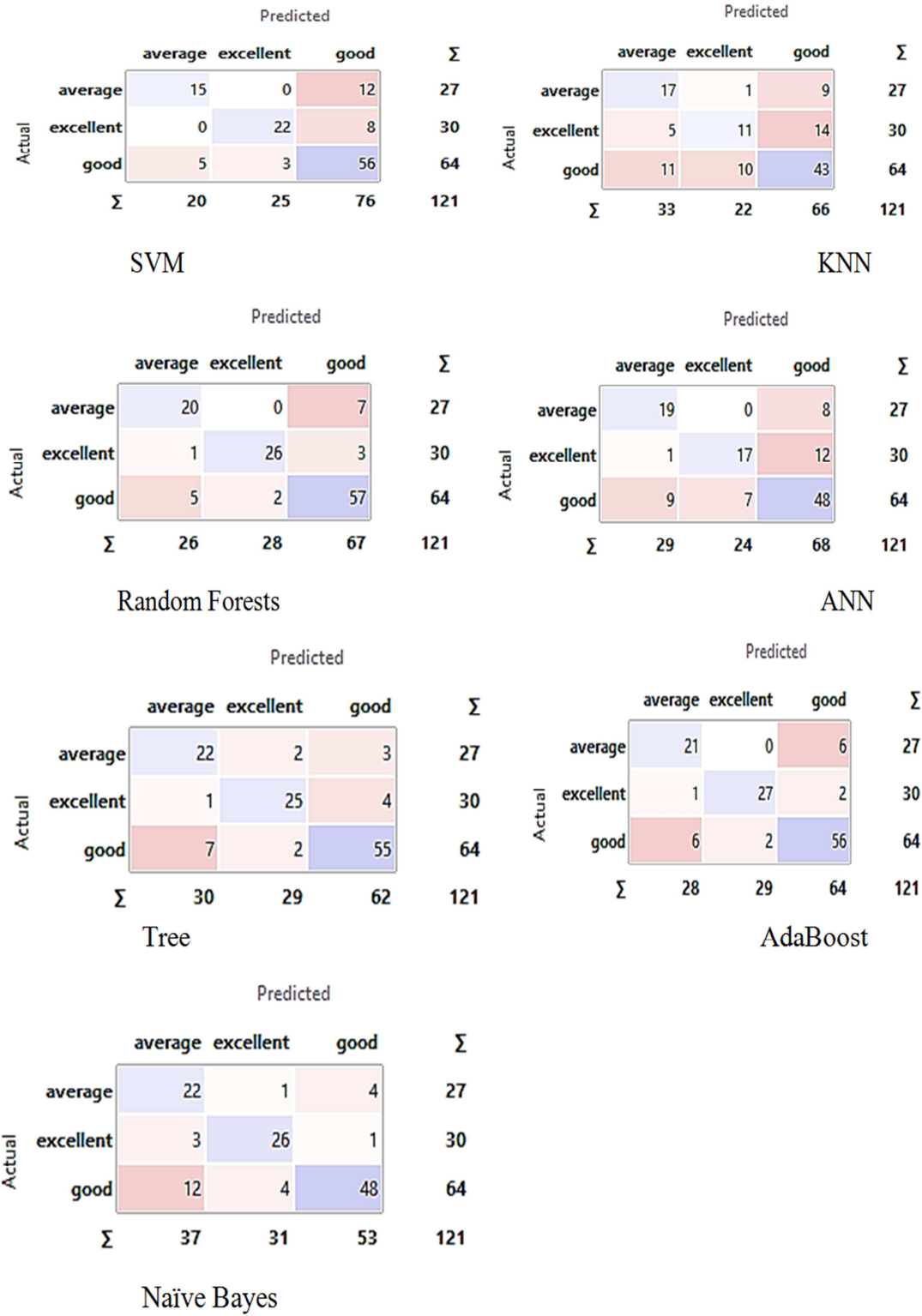

***Figure 2.*** *Confusion matrix for seven algorithms.*

### 3.3. Receiver Operating Characteristic

Receiver operating characteristic is scaled from 0 to 1, 0 means bad prediction and 1 means a high prediction. Seven colors Azura, Burgundy, Emerald, Copper, Pear, Blue-Violet, and Baby blue distinguished to Tree, Neural network, Random forests, KNN, AdaBoost, SVM, and Naïve Bayes respectively. On the X axis there is an FP rate (1 Specificity), and on Y axis there is TP rate (Sensitivity). Figure 3 presents the receiver operating characteristic (ROC) for three instances in the dataset using the output from each applied algorithm.

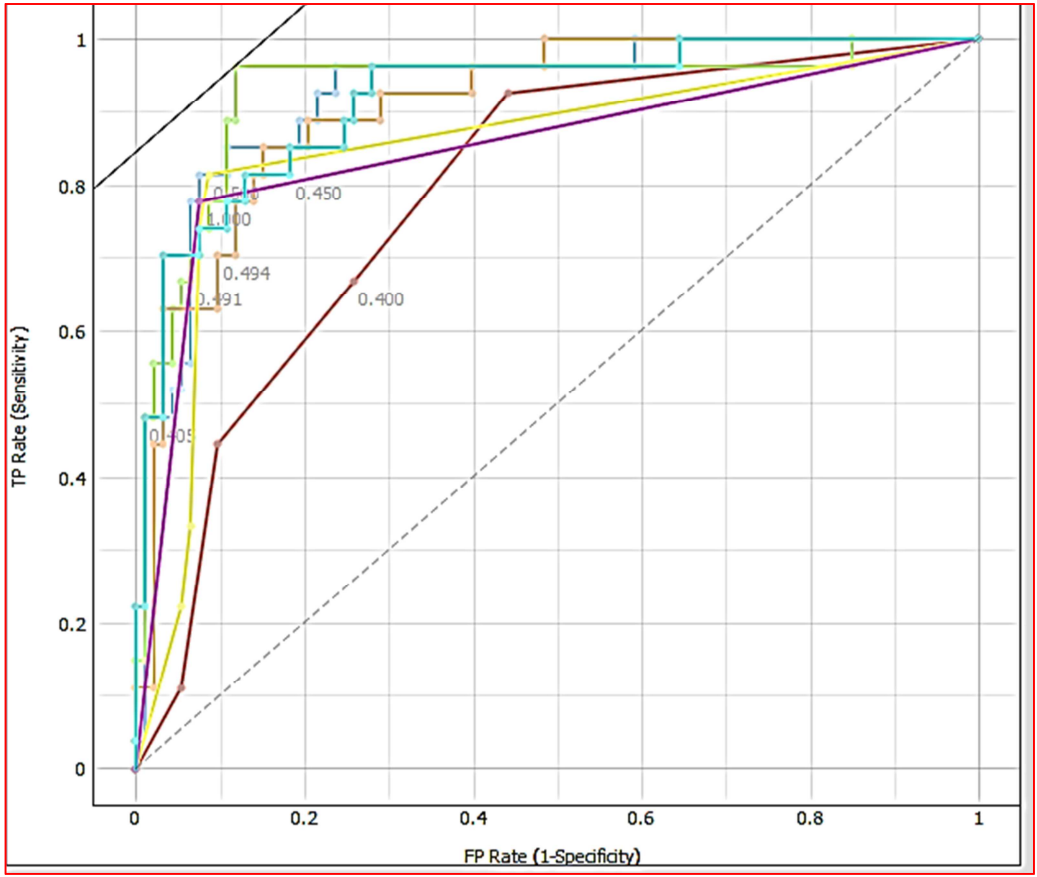

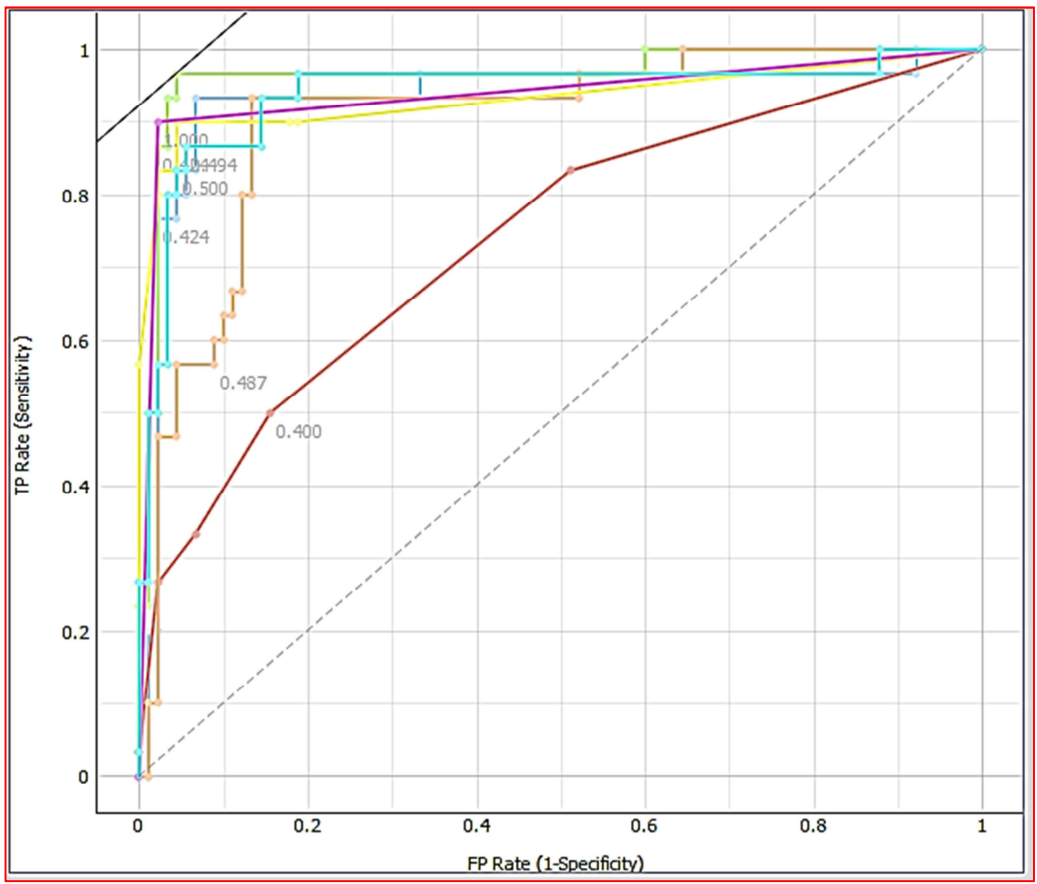

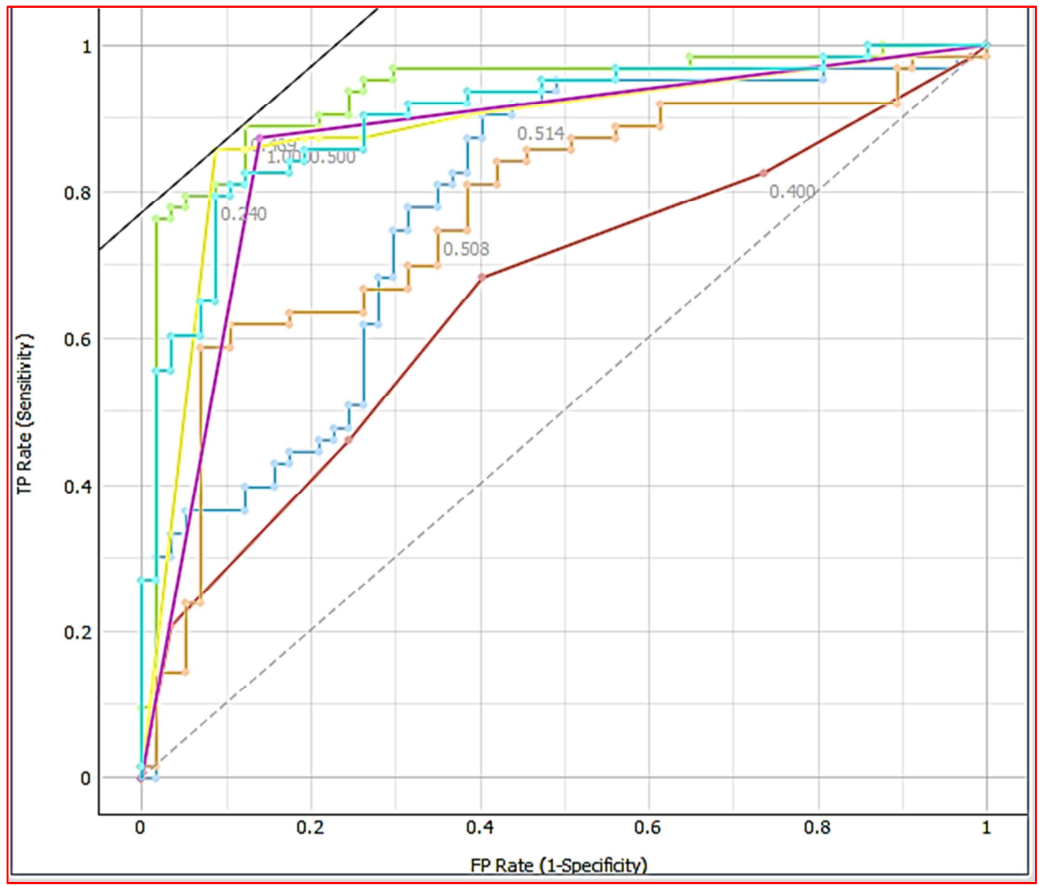

*Figure 3. Receiver operating characteristic for each attribute (ROC curve)*

The algorithm of Random Forests gives the highest area under the curve reaches to (0.932), then Naïve Bayes has (0.908) area under the curve. KNN (K nearest neighbor) recorded the lowest area under the curve among all algorithms reaches to (0.668). Decision tree and its ensemble algorithms recorded the best area under the curve beside Naïve Bayes.

## 4. Conclusion

Traditional analysis and measurements cannot deal with daily data, especially large data. Nowadays, data mining is sufficient to deal with all daily data whether this data is small or huge. In industry, association rules applied successfully to find the relationship between categorical variables, but this algorithm cannot handle numerical data. The study applied seven algorithms to analyze the dataset included categorical and numerical input variables to select the best algorithm can deal with numerical and categorical data. Decision tree and its ensemble algorithms (Random Forests and AdaBoost) recorded the best accuracy and high performance in accuracy classification and the area under the curve (ROC). Therefore. Random Forests, AdaBoost, and Decision tree are appropriate for industrial and manufacturing data. Furthermore, Naïve Bayes gave high accuracy classification and high area under the curve (ROC), but this algorithm cannot deal with numerical data as a decision tree and its ensemble algorithms (Random Forests, and AdaBoost